\documentclass[aps,preprint]{revtex4}%
\usepackage{amsfonts}
\usepackage{amsmath}
\usepackage{amssymb}
\usepackage{graphicx}
\usepackage{color}%
\providecommand{\U}[1]{\protect\rule{.1in}{.1in}}
\begin{document}
\title{Non-Adiabatic Solution to the Time Dependent Quantum Harmonic Oscillator}
\author{C. A. M. de Melo$^{1,2}$\thanks{cassius.anderson@gmail.com}}
\author{B. M. Pimentel$^{2}$\thanks{pimentel@ift.unesp.br}}
\author{J. A. Ramirez$^{2}$\thanks{alrabef@ift.unesp.br} }
\affiliation{$^{1}$Instituto de Ci\^{e}ncia e Tecnologia,}
\affiliation{Universidade Federal de Alfenas, Campus Po\c{c}os de Caldas.}
\affiliation{Rodovia Pref. Jos\'{e} Aur\'{e}lio Vilela (BR 267), Km 533, n$%
%TCIMACRO{\U{b0}}%
%BeginExpansion
{{}^\circ}%
%EndExpansion
$11999, CEP 37715-400, Po\c{c}os de Caldas, MG, Brazil.}
\affiliation{$^{2}$Instituto de F\'{\i}sica Te\'{o}rica, Universidade Estadual Paulista.}
\affiliation{Rua Bento Teobaldo Ferraz 271 Bloco II, P.O. Box 70532-2, CEP 01156-970,
S\~{a}o Paulo, SP, Brazil.}

\begin{abstract}
Using Schwinger Variational Principle we solve the problem of quantum harmonic
oscillator with time dependent frequency. Here, we do not take the usual
approach which implicitly assumes an adiabatic behavior for the frequency.
Instead, we propose a new solution where the frequency only needs continuity
in its first derivative or to have a finite set of removable discontinuities.

\end{abstract}
\maketitle

\section{Introduction}

The study of parametric time-dependent systems like time-dependent quantum
harmonic oscillator is one of the most useful models for Modern Quantum
Mechanics. Its applications cover areas from Quantum Optics to Cosmology.

For instance, in atomic physics, the problem of a charged particle in an
electromagnetic time-dependent field was dealt in the analysis of a Paul Trap
(\cite{Paul, Paul1}, and the references therein).

This kind of device brings a special attention to the simple quantum
mechanical systems . Similar time-dependent problems were investigated by many
authors, like Kulsrud \cite{Kul} in the study of system with adiabatic
behavior of the frequency and Kruskal \cite{Kruskal} who solved the same
problem using canonical transformations of time dependent systems with slowly
changing of the parameters.

Lewis Jr. \cite{Lewis, Lewis1, Lewis2} \textit{et al.} found a kind of
dynamical invariant which leads to deal in a more general form systems like
time-dependent harmonic oscillators. In 1969 Malkin, Man'ko and Trifonov in
\cite{MMT} solved the problem using the theory of dynamical invariants,
proposing a new kind of linear invariants.

Notwithstanding, analytic solutions for the quantum parametric harmonic
oscillator are known only in some sorts of adiabatic regimes. Here, we develop
a new solution which does not need a slow variation of the frequency, opening
a way to new applications in fast varying frequency regimes.

In order to do it we use the Schwinger Quantum Action Principle
\cite{QuanField, QuanField1, QuanField2, QuanField3, QuanField4, QuanField5},
which explores the deep connection between quantum and classical mechanics.
Schwinger Principle has been used recently in several applications as field
theory in curved and torsioned spaces \cite{CQG, CQG1, CQG2}, gauge fixing in
quantum field theories \cite{Bfield} or even to construct a quaternionic
version for the quantum theory \cite{Quaternions}. Such approach have many
advantages both from theoretical or practical points of view. Any model based
in variational principles has deeper fundamental basis and at same time
variational techniques can give rise to new exact and approximated solutions.

For instance, in the case of the quantum harmonic oscillator with variable
frequency, the usual approach is to study the classical problem and to adapt
its solutions to quantum amplitude probability by means an ansatz where the
phase can be varied only adiabatically. However, using a variational approach
we are able to solve the same problem for quickly variations of the frequency.
In order to see how it proceeds let us to start presenting the usual approach.

\section{The conventional ansatz}

The equation of motion for the time dependent harmonic oscillator is
\begin{equation}
\frac{d^{2}q(t)}{dt^{2}}+\omega^{2}(t)q(t)=0.\label{oft0}%
\end{equation}
In \cite{Kul, Lewis} was proposed the following ansatz,%
\begin{equation}
q(t)=S(t)e^{i\gamma(t)},\label{anzatsoft}%
\end{equation}
This give us a solution of Eq. (\ref{oft0}) if, and only if, the differential
equations
\begin{align*}
\ddot{S}(t)+\omega^{2}(t)S(t)  & =\frac{1}{S^{3}(t)},\\
\dot{\gamma}(t)\dddot{\gamma}(t)-\frac{3}{2}\ddot{\gamma}^{2}(t)-2(\omega
^{2}(t)-\dot{\gamma}^{2}(t))\dot{\gamma}^{2}(t)  & =0,
\end{align*}
are satisfied, implying a very soft (adiabatic) behavior for the phase
$\gamma\left(  t\right)  $.

From (\ref{anzatsoft}) one can find the propagator for the quantum system
\cite{khanlaw, fansan}, but the applications are restricted by the adiabatic
assumption \cite{reuter}.

\section{Schwinger Variational Principle}

The Schwinger Variational principle was conceived to settle a reformulation of
Quantum Mechanics without using the correspondence principle. This formulation
establishes that any infinitesimal variation of a transformations function
$\left\langle a\left(  t_{1}\right)  |b\left(  t_{0}\right)  \right\rangle $
can be obtained as the matrix element of a single infinitesimal generator: the
quantum action operator \cite{schwingercd, schwingercd1},%
\begin{equation}
\delta\left\langle a\left(  t_{1}\right)  |b\left(  t_{0}\right)
\right\rangle =i\left\langle a\left(  t_{1}\right)  \right\vert \delta\hat
{S}_{t_{1},t_{0}}\left\vert b\left(  t_{0}\right)  \right\rangle
=i\left\langle a\left(  t_{1}\right)  \right\vert \left.  \left(  \hat
{p}\delta\hat{q}-\hat{H}\delta t\right)  \right\vert _{t_{0}}^{t_{1}%
}\left\vert b\left(  t_{0}\right)  \right\rangle ,\label{eq:actprinc}%
\end{equation}
where $\delta\hat{S}_{t_{1},t_{0}}=\delta\left[  \hat{S}_{t_{1},t_{0}}\right]
$ and $\hat{S}_{t_{1},t_{0}}=\int_{t_{0}}^{t_{1}}\hat{L}(t)dt$.

One essential apparatus of this formalism is the generator $\hat{G}$ defined
by%
\[
\hat{G}=\hat{p}\delta\hat{q}-\hat{H}\delta t.
\]
Fixing the boundary conditions on the states in (\ref{eq:actprinc}), leads to
the Schr\"{o}dinger equation and fixing the boundary conditions on the
operators results in the Heisenberg picture.

In order to solve (\ref{eq:actprinc}) and to obtain the transformation
function, it is necessary to order the action function such that,%
\[
\delta\langle a(t_{1})|b(t_{0})\rangle=i\langle a(t_{1})|\delta\hat{S}%
_{t_{1},t_{0}}|b(t_{0})\rangle=i\delta\mathcal{W}_{t_{1},t_{0}}\langle
a(t_{1})|b(t_{0})\rangle
\]
obtaining,%
\begin{equation}
\langle a(t_{1})|b(t_{0})\rangle=e^{i\mathcal{W}_{t_{0},t_{1}}}%
.\label{amplitude}%
\end{equation}

\section{The Non-Adiabatic Solution}

Instead of using (\ref{anzatsoft}) we propose a new solution given by%
\begin{equation}
q(t)=A(t)\exp\left[  \pm i\int_{t_{0}}^{t}\omega(\tau)d\tau\right]
\label{anzatsoft1}%
\end{equation}
which imply a second order differential equation
\begin{equation}
\ddot{A}(t)+2i\omega(t)\dot{A}(t)+i\dot{\omega}(t)A(t)=0,\label{oft3}%
\end{equation}
for the amplitude. Here, one needs only a frequency function $\omega(t)$ with
\emph{first derivative continuous} or having a finite set of removable discontinuities.

On this way, the general solution for the quantum analog problem given in
(\ref{oft0}) is
\begin{equation}
\hat{q}=\frac{\hat{q}_{0}}{C(t_{0})}F_{1}(t)+\frac{\hat{p}_{0}}{mC(t_{0}%
)}F_{0}(t),\label{oft5.1}%
\end{equation}
where
\begin{align*}
F_{0}(t)  &  =A_{0}A^{\ast}(t)\exp\left[  -i\int_{t_{0}}^{t}\omega(\tau
)d\tau\right]  -c.c.,\\
F_{1}(t)  &  =\left[  \dot{A}_{0}^{\ast}-i\omega_{0}A_{0}^{\ast}\right]
A(t)\exp\left[  i\int_{t_{0}}^{t}\omega(\tau)d\tau\right]  -c.c.,\\
C\left(  t\right)   &  =A^{\ast}(t)\dot{A}(t)-A(t)\dot{A}^{\ast}%
(t)+2i\omega(t)\left\vert A(t)\right\vert ^{2}%
\end{align*}

\section{Quantum Transition Amplitude}

Taking the non-adiabatic solution for $\hat{q}(t)$ we can find the canonical
momentum
\begin{equation}
\hat{p}=-m\hat{q}_{0}\frac{C(t)}{F_{0}(t)}+\hat{q}m\frac{\dot{F}_{0}(t)}%
{F_{0}(t)},\label{momentum}%
\end{equation}

The commutator of $\hat{q}$ in different times can be reached using the
canonical relation $\left[  \hat{q},\hat{p}\right]  =i\hbar$,%

\begin{equation}
\hat{q}_{0}\hat{q}=\frac{i\hbar F_{0}(t)}{mC(t)}+\hat{q}\hat{q}_{0}%
.\label{relcom}%
\end{equation}

The quantum Hamiltonian has the form:%

\begin{align*}
\hat{H} & =\frac{\hat{p}^{2}}{2m}+\frac{1}{2}m\omega^{2}(t)\hat{q}^{2}=\\
& =\frac{m}{2}\left[  \hat{q}^{2}\frac{\dot{F}_{0}^{2}\left(  t\right) }%
{F_{0}^{2}\left(  t\right)  }+\hat{q}_{0}^{2}\frac{C^{2}\left(  t\right)
}{F_{0}^{2}\left(  t\right)  }-2\frac{C\left(  t\right)  \dot{F}_{0}\left(
t\right)  }{F_{0}^{2}\left(  t\right)  }\hat{q}\hat{q}_{0}\right]
-\frac{i\hbar}{2}\frac{\dot{F}_{0}\left(  t\right)  }{F_{0}\left(  t\right)
}+\frac{m}{2}\omega^{2}\left(  t\right)  \hat{q}^{2},
\end{align*}

Therefore, the transition amplitude $\left\langle q,t|q_{0},t_{0}\right\rangle
$ in the Schwinger formulation is given for
\begin{align*}
\left\langle q,t|q_{0},t_{0}\right\rangle  &  =\mathcal{A}\left(
q,q_{0}\right)  \exp\left\{  -\frac{im}{2\hbar}\int_{t_{0}}^{t}H\left(
t\right)  dt\right\}  =\\
&  =\mathcal{A}\left(  q,q_{0}\right)  \exp\left\{  \frac{im}{2\hbar
F_{0}\left(  t\right)  }\left(  q^{2}\dot{F}_{0}(t)+q_{0}^{2}F_{1}%
(t)-qq_{0}C(t)\right)  \right\}
\end{align*}

where one can recognize the classical action%

\[
S(q,q_{0},t)=\frac{m}{2F_{0}(t)}\left(  q^{2}\dot{F}_{0}(t)+q_{0}^{2}%
F_{1}(t)-qq_{0}C(t)\right)  .\label{tranamp}%
\]

We can verify this expression, recovering the action for the harmonic
oscillator with constant frequency taking the regime $\omega(t)\rightarrow
\omega_{0}$, and the amplitudes $A(t)$ constants, then%

\[
\underset{\omega(t)\rightarrow\omega_{0}}{\lim}S(q,q_{0},t)=\frac{m}%
{2\sin\left[ \omega_{0}(t-t_{0})\right] }\left( \left\{  q^{2}+q_{0}%
^{2}\right\}  \cos\left[ \omega_{0}(t-t_{0})\right] -qq_{0}\right) .
\]

In order to fix the explicit form of $\mathcal{A}\left( q,q_{0}\right) $ we use%

\[
\frac{\partial\left\langle q,t|q_{0},t_{0}\right\rangle }{\partial q}=\frac
{i}{\hbar}\left\langle q,t\right\vert \hat{p}\left\vert q_{0},t_{0}%
\right\rangle ,
\]

obtaining%

\begin{align}
\frac{\partial\left\langle q,t|q_{0},t_{0}\right\rangle }{\partial q} =\left(
\frac{1}{\mathcal{A}\left( q,q_{0}\right) }\frac{\partial\mathcal{A}\left(
q,q_{0}\right) }{\partial q}+\frac{i}{\hbar}qm\frac{\dot{F}_{0}(t)}{F_{0}%
(t)}-\frac{i}{\hbar}q_{0}m\frac{C(t)}{F_{0}(t)}\right) \left\langle
q,t|q_{0},t_{0}\right\rangle .
\end{align}

and comparing with the expressions for momentum $\hat{p}$ and position
$\hat{q}$ we obtain in booth of cases%

\[
\frac{\partial\mathcal{A}\left(  q,q_{0}\right)  }{\partial q}=\frac
{\partial\mathcal{A}\left(  q,q_{0}\right)  }{\partial q_{0}}=0,
\]

then the function $\mathcal{A}\left( q,q_{0}\right)  = \mathcal{A}\left(
t\right) $ only depends on time and%

\[
\left\langle q,t|q_{0},t_{0}\right\rangle =\frac{\mathcal{A}\left(  t\right)
}{\sqrt{F_{0}(t)}}\exp\left\{  \frac{i}{\hbar}\frac{m}{2F_{0}(t)}\left(
q^{2}\dot{F}_{0}(t)+q_{0}^{2}F_{1}(t)-2qq_{0}C(t)\right)  \right\}  ,
\]
and the explicit form of $\mathcal{A}$ can be obtained from $\lim
_{t_{0}\rightarrow t_{1}}\left\langle q|q_{0}\right\rangle =\delta\left(
q-q_{0}\right) $, comparing with a suitable sequence of functions converging
to the Dirac delta, like the following%

\[
\lim_{t_{0}\rightarrow t_{1}}\left\langle q,t|q_{0},t_{0}\right\rangle =
\lim_{t_{0}\rightarrow t_{1}}\frac{K}{\sqrt{F_{0}(t)}}\exp\left\{  -\frac
{i}{\hbar}\frac{mC(t_{0})}{2F_{0}(t)}\left( q-q_{0}\right) ^{2}\right\} .
\]

then one finds $\mathcal{A}\left(  t\right)  = \sqrt{\frac{imC(t)}{2\pi\hbar}%
}$, and finally the final form for the transformation function:
\[
\left\langle q,t|q_{0},t_{0}\right\rangle =\sqrt{\frac{imC(t)}{2\pi\hbar
F_{0}(t)}}\exp\left\{  \frac{i}{\hbar}\frac{m}{2F_{0}(t)}\left(  q^{2}\dot
{F}_{0}(t)+q_{0}^{2}F_{1}(t)-2qq_{0}C(t)\right)  \right\}  ,
\]

\section{Conclusions and Perspectives}

The last result can be applied in many areas as the study of electromagnetic
cavities \cite{Dodonov, Dodonov1, Dodonov2} or any other quantum particle
interacting with a classical variable harmonic potential. When we deal with a
more realistic behavior for the external fields it is common to find a
non-adiabatic variation of its parameters. Therefore, the presented solution
provides a more accurate description in this case, with the advantage of to
reproduce the know results for adiabatic regimes.

\section{Aknowledgments}

C.A.M. de Melo thanks FAPEMIG for partial support, B.M.Pimentel thanks CNPq
for partial support and J.A.Ramirez thanks CAPES for full support.


\begin{thebibliography}{99}                                                                                               %
\bibitem {Paul}W. Paul, H. Steinwedel, \textit{Zeitschrift f\"{u}r
Naturforschung A} \textbf{8} (1953) 448-450.

\bibitem {Paul1}Wolfgang Paul, \textit{Rev. Mod. Phys} \textbf{62}, (1990) 531.

\bibitem {Kul}Russell M. Kulsrud, \textit{Phys.Rev.} \textbf{106}, (1957) 205.

\bibitem {Kruskal}Martin Kruskal, \textit{Jour. Math. Phys.} \textbf{3},
(1962) 4.

\bibitem {Lewis}H. R. Lewis, Jr. , \textit{Phys. Rev. Lett.} \textbf{18},
(1967) 510.

\bibitem {Lewis1}H. R. Lewis, Jr. , \textit{Jour. Math. Phys.} \textbf{9},
(1968) 1976.

\bibitem {Lewis2}H. R. Lewis, Jr. and W. B. Riessenfeld , \textit{Jour. Math.
Phys.} \textbf{10}, (1969) 1458.

\bibitem {MMT}I. A. Malkin, V. I. Man'ko and D. A. Trifonov. , \textit{Phys.
Lett.} \textbf{30A}, (1969) 7.

\bibitem {QuanField}J. S. Schwinger, \textit{Phys. Rev.} \textbf{82}, (1951) 914.

\bibitem {QuanField1}J. S. Schwinger, \textit{Phys. Rev.} \textbf{91}, (1953) 713.

\bibitem {QuanField2}J. S. Schwinger, \textit{Phys. Rev.} \textbf{91}, (1953) 728.

\bibitem {QuanField3}J. S. Schwinger, \textit{Phys. Rev.} \textbf{92}, (1953) 1283.

\bibitem {QuanField4}J. S. Schwinger, \textit{Phys. Rev.} \textbf{93}, (1954) 615.

\bibitem {QuanField5}J. S. Schwinger, \textit{Phys. Rev.} \textbf{94}, (1954) 1362.

\bibitem {CQG}R. Casana, C. A. de Melo and B. M. Pimentel, \emph{Class. Quant.
Grav.}.\textbf{24}, (2007) 723.

\bibitem {CQG1}R. Casana, C. A. de Melo and B. M. Pimentel, \textit{Astrophys.
Sp. Sci.} \textbf{305}, (2006) 125.

\bibitem {CQG2}R. Casana, C. A. de Melo and B. M. Pimentel, \textit{Braz. J.
Phys.} \textbf{35}, (2005) 1151.

\bibitem {Bfield}C. A. M. de Melo, B. M. Pimentel and P. J. Pompeia,
\textit{Il Nuovo Cim.} \textbf{B121}, (2006) 193.

\bibitem {Quaternions}C. A. M. de Melo and B. M. Pimentel, \textit{Adv. App.
Clif. Alg. }DOI: 10.1007/s00006-010-0234-8 (2010). Available online.

\bibitem {khanlaw}D. C. Khandekar and S. V. Lawande, \textit{Jour. Math.
Phys.} \textbf{16} (1975) 384.

\bibitem {fansan}C. Farina and A. J. S. Santoja, \textit{Phys. Lett.}
\textbf{A184 }(1993) 23.

\bibitem {reuter}W. Dietrich and M. Reuter, \emph{Classical and Quantum
Dynamics: From Classiclal paths to Path Integrals}, Chap. 8, 3rd
edition{\ }(Springer-Verlag, 2001).

\bibitem {schwingercd}J. S. Schwinger, \textit{\emph{Quantum Kinematics and
Dynamics}}, (W.A. Benjamin Publishers, 1970);

\bibitem {schwingercd1}J. S. Schwinger, \emph{Quantum Mechanics: Symbolism of
Atomic Measurements} (Springer, 2001).

\bibitem {Dodonov}V. V. Dodonov, A. B. Klimov and D. E. Nikonov, \textit{Phys.
Rev} \textbf{A47} (1993) 4442.

\bibitem {Dodonov1}V. V. Dodonov, V. I. Man'ko and D. E. Nikonov,
\textit{Phys. Rev.} \textbf{A51} (1995) 3328.

\bibitem {Dodonov2}V.V. Dodonov and A.V. Dodonov, \textit{J. Rus. Laser
Research} \textbf{26} (2005) 6.
\end{thebibliography}
\end{document}